\documentclass[12pt]{iopart}
\usepackage{graphicx}
\usepackage{iopams}
\expandafter\let\csname equation*\endcsname\relax
\expandafter\let\csname endequation*\endcsname\relax
\usepackage{amsmath}
\usepackage{amssymb}
\usepackage{csquotes}
\MakeOuterQuote{"}
\usepackage{bm}

\begin{document}

\title{Compressional Alfv\'en eigenmodes excited by runaway electrons}

\author{Chang Liu}
\address{Princeton Plasma Physics Laboratory, Princeton, NJ, United States of America}
\ead{cliu@pppl.gov}
\author{Dylan P. Brennan}
\address{Princeton University, Princeton, NJ, United States of America}
\author{Andrey Lvovskiy}
\address{Oak Ridge Associated Universities, Oak Ridge, TN, United States of America}
\author{Carlos Paz-Soldan}
\address{General Atomics, San Diego, CA, United States of America}
\author{Eric D. Fredrickson}
\address{Princeton Plasma Physics Laboratory, Princeton, NJ, United States of America}
\author{Amitava~Bhattacharjee}
\address{Princeton Plasma Physics Laboratory, Princeton, NJ, United States of America}
\address{Princeton University, Princeton, NJ, United States of America}

\begin{abstract}
	
Compressional Alfv\'en eigenmodes (CAE) driven by energetic ions have been observed in magnetic fusion experiments. In this paper, we show that the modes can also be driven by runaway electrons formed in post-disruption plasma, which may explain kinetic instabilities observed in DIII-D disruption experiments with massive gas injection. The mode-structure is calculated, as are the  frequencies which are in agreement with experimental observations. Using a runaway electron distribution function obtained from a kinetic simulation, the mode growth rates are calculated and found to exceed the collisional damping rate when the runaway electron density exceeds a threshold value. The excitation of CAE poses a new possible approach to mitigate seed runaway electrons during the current quench and surpassing the avalanche.

\end{abstract}


\maketitle

\section{Introduction}

The generation of high-energy runaway electrons poses one of the major threats to the safety and reliable performance of tokamak fusion reactors \cite{lehnen_disruptions_2015}. It is predicted that in a typical unmitigated disruption in ITER, a large fraction of plasma current can be converted to be carried by runaway electrons\cite{boozer_pivotal_2018}, which can be accelerated to tens of MeV and cause severe damage to the plasma facing material. Experimental and theoretical studies have been conducted to search effective mitigation strategies. Injection of a large amount of impurities into the post-disruption plasma is considered to be the basic mitigation strategy for ITER\cite{lehnen_disruptions_2015,hollmann_measurement_2015}. However, it is unclear whether the additional collisional damping brought by impurities is enough to suppress the generation of runaway electrons near the core, given the fast multiplication of RE population (avalanche) due to secondary generation. Therefore alternative mitigation methods are also under investigation, including magnetic perturbations introduced by external magnetic coils\cite{chen_suppression_2018}, and wave-particle interactions introduced by either kinetic instabilities\cite{liu_role_2018} or injection of radio-frequency (RF) waves \cite{guo_control_2018}.

The importance of kinetic instabilities associated with runaway electrons have been shown in both experiments\cite{paz-soldan_spatiotemporal_2017,lvovskiy_observation_2019} and numerical simulations\cite{liu_role_2018}. Fan instabilities\cite{parail_kinetic_1978}, which are driven by the anisotropic distribution of runaway electrons, can lead to enhanced pitch angle scattering of resonant electrons. Whistler waves produced by fan instabilities have been directly observed in the DIII-D experiments with runaway electrons generated during the quiescent phase\cite{spong_first_2018}. In addition, these instabilities can significantly enhance the power of electron cyclotron emission (ECE) signals from runaway electrons\cite{liu_effects_2018}, and help suppress RE avalanche and increase the critical electric field\cite{paz-soldan_growth_2014,liu_role_2018}. Although whistler waves can have strong resonance with runaway electrons, they will be susceptible to strong collisional damping in a disruption scenario. The collisional damping of the waves, which is similar to a resistive damping in a electromagnetic oscillation circuit, can lead to a very high threshold for the modes to occur\cite{aleynikov_stability_2015}. In addition, the collisional damping can be stronger in the case with impurity injections, which can introduce more cold electrons as colliders.

Recently a new kind of instability has been observed in the post-disruption plasmas in DIII-D experiments\cite{lvovskiy_role_2018}. This instability is used to explain the RE loss during current quench and the failure of RE plateau formation, which has been observed with argon massive gas injection (MGI) in relatively small amounts. The magnetic perturbations of these instabilities are measured using the ion cyclotron emission (ICE) coils. The frequency of the mode associated is in the range of 0.5-3 MHz, which is much lower compared to that of whistler waves, and the physical mechanism underlying the mode is not well understood. Nevertheless, given the strong correlation between the occurrence of such instabilities and the loss of runaway electrons, the ability to drive the instabilities provides a new and promising approach to mitigate the seed runaway electrons before a significant avalanche happens.

In this paper, we give an explanation of the observed kinetic instabilities in the post-disruption plasma, which is the compressional Alfv\'en eigenmode (CAE). The CAE is the high frequency branch (fast wave) of Alfv\'en waves with magnetic perturbation in both the parallel ($B_\parallel$) and the perpendicular ($B_\perp$) directions with respect to the equilibrium magnetic field. CAEs driven by energetic ions from neutral beam injection (NBI) have been observed in spherical toruses (STs) like NSTX\cite{fredrickson_non-linear_2013} and MAST\cite{smith_compressional_2017}. Compared to whistler waves, CAEs are less susceptible to collisional damping and easier to excite in disruption plasmas. However, it is  more difficult to satisfy the resonance condition with runaway electrons whose gyrofrequencies and transit frequencies are much higher compared to the mode frequency. Here we find that for trapped and barely passing runaway electrons, the resonance condition can be satisfied, and the gradient of RE distribution function in both momentum space and in the radial direction can drive the mode to become unstable. A similar mechanism has been used to explain the excitation of beta-induced Alfv\'en eigenmodes (BAE) by high energy electrons observed in the HL-2A tokamak\cite{hl-2a_team_$ensuremathbeta$-induced_2010}. In addition, the injection of high-Z impurities can help the generation of trapped and barely passing runaway electrons through the partial screening effect\cite{hesslow_effect_2017} and the excitation of instabilities. Using the plasma parameters from the DIII-D experiments, we can calculate the frequencies and mode structures of CAEs in the post-disruption scenario, and the mode growth rate using RE distribution function from a kinetic simulation, Our results appear to be consistent with experimental observations.

This paper is organized as follows. In Sec.~\ref{sec:mode-structure} the frequency and the mode structure of CAEs in a disruptive plasma is calculated using a simplified model for the  CAE dispersion relation. In Sec.~\ref{sec:kinetic-model} a kinetic model for simulating the RE distribution function $f$ is introduced, which is implemented in the code QUADRE. This model introduces both the enhanced pitch angle scattering due to partial screening of high-$Z$ impurities, and a bounce-average model for both passing and trapped electrons. In Sec.~\ref{sec:wave-particle} we show the calculation of mode growth rate due to RE gradient in both momentum space and in the radial direction through energy exchange. In Sec.~\ref{sec:collisional-damping} the collisional damping of CAE, which plays a determining role for mode stability, is calculating using a two fluid model. In Sec.~\ref{sec:linear-simulation} the result of a linear simulation is presented, including the evolution of the RE distribution function and results of mode growth rates. In Sec.~\ref{sec:discussion} the results are discussed including the possible feedback effect of CAEs on REs.

\section{Frequencies and eigenmodes of compressional Alfv\'en eigenmodes}
\label{sec:mode-structure}

In this section we show the calculation of CAE eigenmode frequencies and mode structures based on the plasma parameters from the experiments. The calculation utilizes a simplified Alfv\'en wave dispersion relation. The frequencies of excited modes observed in DIII-D experiments are in the range of $0.5$~MHz to $3$~MHz, which is in the range of Global Alfv\'en Eigenmodes (GAE) or CAE. In this paper, we will focus on the the modeling of CAE and try to use that to explain the experiments. The possibility of GAE will be discussed in Sec.~\ref{sec:discussion}.

For CAE, we first focus on the low frequency modes observed in experiments with frequencies $\omega\ll\omega_{ci}$. In this case, the dispersion relation of CAE can be regarded as isotropic, which can simplify the mode calculation. By introducing the toroidal mode number $n$, the CAE dispersion relation in a torus can be simplified as
\begin{equation}
\label{eq:laplacian}
\nabla_\mathrm{pol} \cdot v_A^2 B_0^2\nabla_{\mathrm{pol}} b_\parallel =(\frac{n^2}{R^2}v_A^2-\omega^2) B_0^2 b_\parallel,
\end{equation}
where $v_A^2=B_0^2/4\pi n_i m_i$ represents the local Alfv\'en velocity ($n_i=n_e/Z$ is the ion density, $n_e$ is the electron density, $Z$ is the ion charge, and $m_i$ is the ion mass), $R$ is the major radius, $\omega$ is the mode frequency, $b_\parallel=\tilde{B}_\parallel/B_0$, where $\tilde{B}_\parallel$ is the mode magnetic field parallel to the equilibrium $B_0$ field, and $\nabla_{\mathrm{pol}}$ is gradient operator projected on the poloidal plane. Note that this simplified dispersion relation is not directly derived from the MHD equations, but obtained by combining the MHD equations with the CAE dispersion relation $\omega^2=(k_\parallel^2+k_\perp^2) v_A^2$ and assuming $k_\perp\gg k_\parallel$. The details of the derivation are given in Ref.~\cite{smith_compressional_2017}.

This dispersion relation is similar to Eq.~(2) in Ref.~\cite{fredrickson_non-linear_2013}. However, Eq.~(2) in Ref.~\cite{fredrickson_non-linear_2013} is a differential equation for the parallel electric field, and a Dirichlet boundary condition is applied. In our case, we solve the equation for $\tilde{B}_\parallel$, so a Neumann boundary condition should be used instead to satisfy the restriction of the magnetic fields at a conducting wall ($\partial B/\partial n=0$).

It is not trivial to solve both the eigenvalue and eigenvectors of Eq.~(\ref{eq:laplacian}) in a 2D mesh.
Here we use a simple numerical method, which is to convert Eq.~(\ref{eq:laplacian}) into a diffusion equation by adding a time derivative term,
\begin{equation}
\frac{1}{B_0^2}\nabla_\mathrm{pol} \cdot v_A^2 B_0^2\nabla_{\mathrm{pol}} b_\parallel+\left(\omega^2-\frac{n^2}{R^2}v_A^2\right)b_\parallel=\frac{\partial}{\partial \tau} b_\parallel.
\end{equation}
We can then solve the equation as a time-dependent problem by giving an initial guess of $\omega$ and the eigenfunction $b_\parallel$. At each timestep, $\omega$ is adjusted as
\begin{equation}
\omega^2=-\frac{\langle \left[ (1/B_0^2)\nabla_\mathrm{pol} \cdot v_A^2 B_0^2\nabla_{\mathrm{pol}}-n^2v_A^2/R^2\right]b_\parallel|b_\parallel\rangle}{\langle b_\parallel|b_\parallel\rangle},
\end{equation}
where $\langle\cdots|\cdots\rangle$ is the inner product and is calculated by integrating the product of two functions in the poloidal plane.
In addition, the Neumann boundary condition is applied to the newly obtained solution of $b_\parallel$ at each timestep. In solving this diffusion equation, part of the guess function which does not satisfy the dispersion relation will decay very quickly, and $\omega$ will converge to the lowest frequency mode. After obtaining the eigenfunction corresponding to the lowest frequency mode, one can repeat this process and find the next higher frequency  eigenmode by only keeping the part of $b_\parallel$ that is orthogonal to the  previously found eigenfunctions.

\begin{figure}[h]
	\begin{center}
		\includegraphics[width=0.2\linewidth]{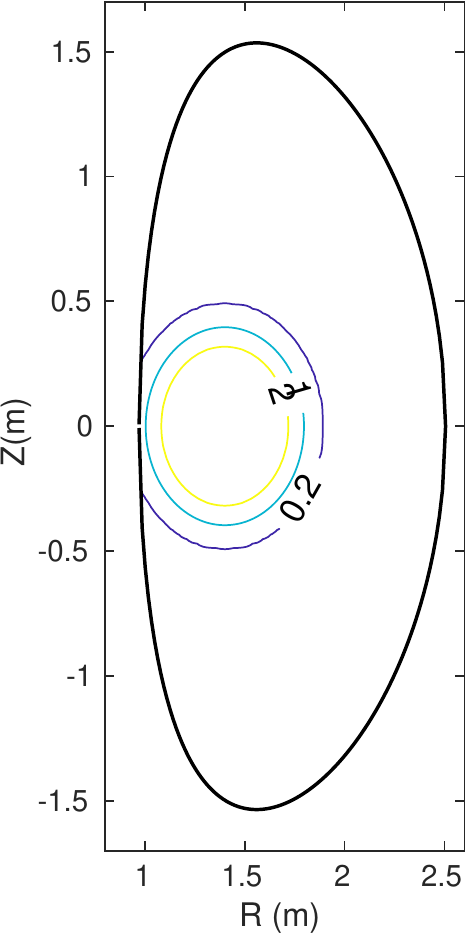}
		\includegraphics[width=0.2\linewidth]{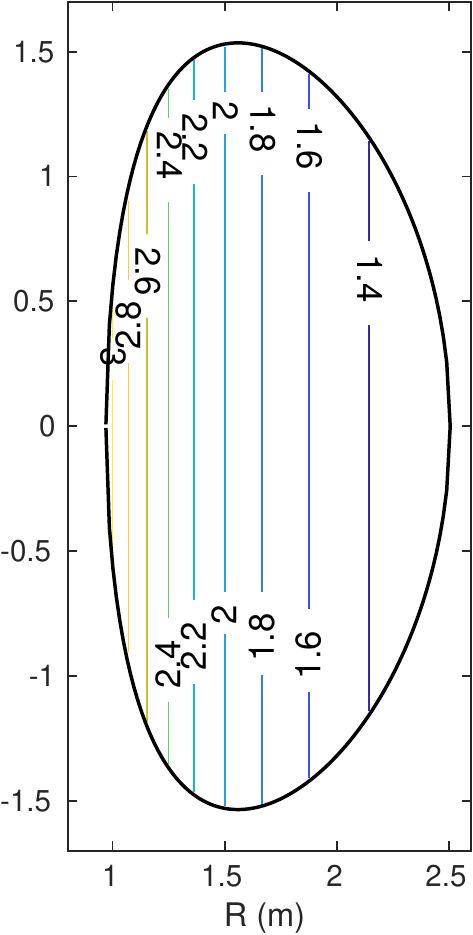}
	\end{center}
	\caption{\label{fig:density-profile}Density (left, unit $10^{20}$~m$^{-3}$) and magnetic field $B_0$ (right, unit T) structure used in solving eigenmodes.}
\end{figure}

The plasma parameters used in solving the eigenmodes correspond to the DIII-D disruption experiments. The density and magnetic field structure in the poloidal plane are shown in Fig.~\ref{fig:density-profile}. The plasma density is strongly localized near the high field side with a peak value $2\times 10^{20}$~m$^{-3}$, due to the shrinking of the current channel being scraped off at the wall in the current quench, which has been confirmed in the experiment using EFIT\cite{paz-soldan_kink_2019}. The vertical instabilities are suppressed in this case by external control coils. In the region outside the last closed flux surface, we set a nonzero plasma density with $n=0.1n_{\mathrm{core}}$, reflecting the weak ionization of the injected gas. This halo plasma can reduce the decay of the mode in the evanescent region. We assume that the plasma is composed of Ar with $Z=2$, and the magnetic field at axis is 2~T. In this case, the Alfv\'en velocity $v_A$ at the core is about $6.9\times 10^5$~m/s.

\begin{figure}[h]
	\begin{center}
		\includegraphics[width=0.7\linewidth]{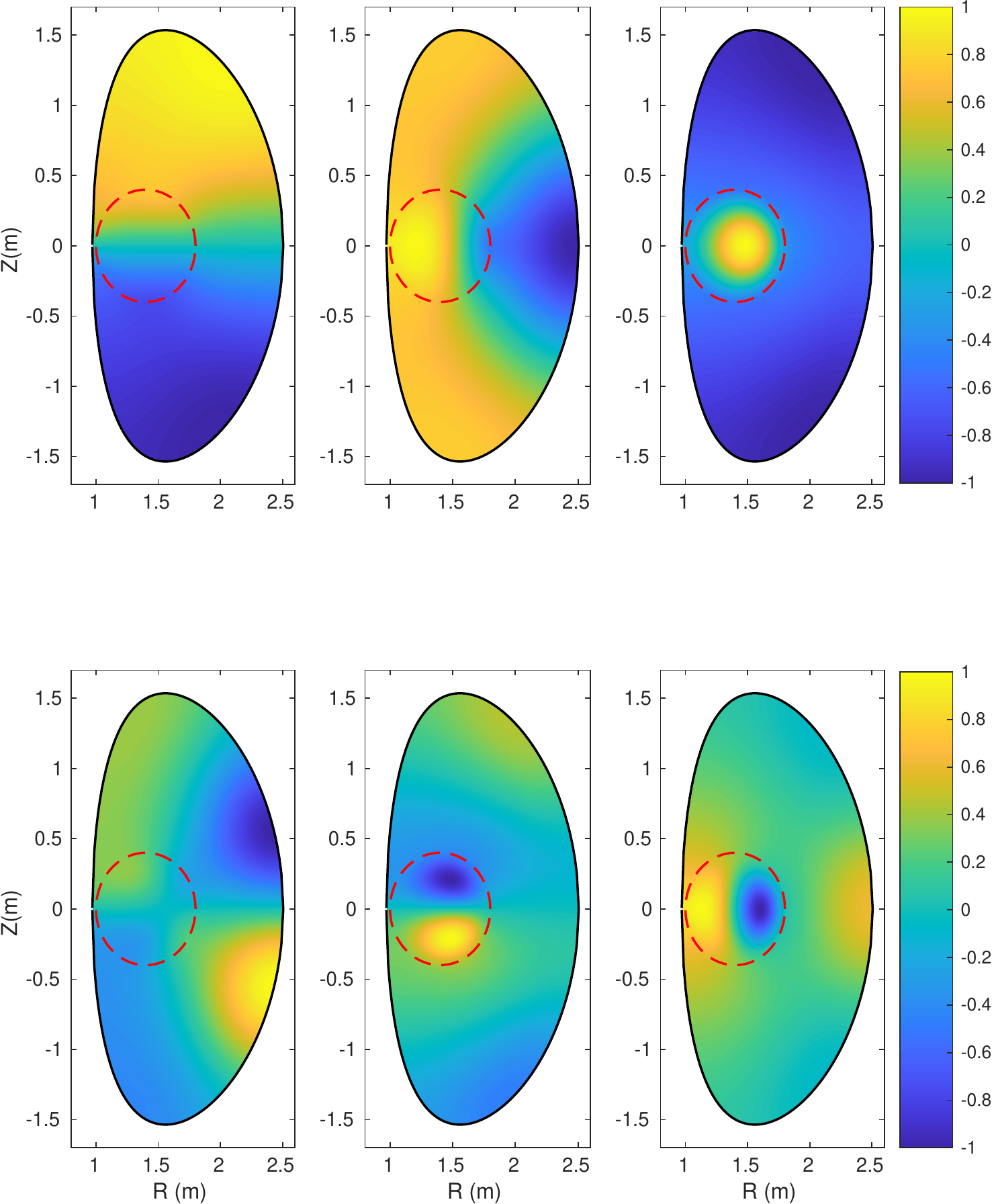}
	\end{center}
	\caption{\label{fig:mode-structure}The mode structure of six lowest frequency CAE with $n=1$ calculated using Eq.~(\ref{eq:laplacian}). The red dash line shows the location of high density plasma. The corresponding frequencies are 0.43~MHz, 0.65~MHz, 0.72~MHz, 0.93~MHz, 1.03~MHz, and 1.04~MHz.}
\end{figure}

Fig.~\ref{fig:mode-structure} shows mode structure ($b_\parallel$) of the six lowest frequency eigenmodes with $n=1$ calculated from Eq.~(\ref{eq:laplacian}). The corresponding eigenfrequencies are $\omega_1=0.43$~MHz, 0.65~MHz, 0.72~MHz, 0.93~MHz, 1.03~MHz, and 1.04~MHz. These frequencies are consistent with the frequencies in the CAE spectrogram observed in experiments\cite{lvovskiy_role_2018}. The red dashed lines show the location of high density plasma corresponding to Fig.~\ref{fig:density-profile}. Judging from the mode structure inside this region, mode 1,2,5,6 can be regarded as $m=1$, mode 3 can be seen as $m=0$, and mode 4 can be seen as $m=2$. Using the eigenmode solver, we can further calculate the mode frequency of larger value of $m$ for $n$. However, for high frequency modes with $\omega\approx \omega_{ci}$, the dispersion relation for CAE becomes anisotropic, and Eq.~(\ref{eq:laplacian}) cannot be be used to calculate the eigenmodes. Nevertheless, this branch of mode (fast wave branch) where CAE persists can extend to high frequencies with $\omega>\omega_{ci}$, which is consistent with experimental observations\cite{lvovskiy_role_2018}.

Note that the frequencies of the Alfv\'en eigenmodes are proportional to $v_A/r$, where $v_A$ depends on the local values of  $n_e$, $B$ and $Z$. Therefore, for a different tokamak with a different plasma condition, one can estimate the eigenmode frequency by multiplying the ratio of $v_A$ with respect to the value used here. In addition, according to the CAE dispersion relation, if decreasing the ion charge of argon $Z$ from $2$ to $1$, the eigenfrequencies of the same mode can drop by a factor of $\sqrt{2}$ given unchanged $n_e$. This may explain the frequency drop of excited modes observed in DIII-D experiments\cite{lvovskiy_role_2018}, where the ion density is increasing due to the continuing gas puffing, but ion charge number $Z$ is decreasing due to the dropping plasma temperature.
  
\section{Kinetic model of runaway electrons}
\label{sec:kinetic-model}

The seed REs in disruption experiments can be generated through the hot-tail mechanism\cite{smith_hot_2008}, which occurs due to the Maxwellian tail of the thermal electrons before the thermal quench. These seed electrons can then be dragged into higher energy regime by the inductive electric field and become runaway electrons, thus providing the free energy for kinetic instabilities such as CAEs. In this study we used a simplified model to simulate hot-tail generation of REs with a prescribed electric field which is initially large enough to drive the hot tail, and then drop to a fixed value which corresponds to the loop voltage in experiments after the tail is generated.

We use the QUADRE code to simulate the generation and dynamics of REs this process. QUADRE uses a continuum description for electron distribution in momentum space, which includes both the Maxwellian part and the runaway tail. The relativistic effect of high energy electrons are taken into account. The collision of both bulk and runaway electrons, including the slowing-down and pitch angle scattering are described using a unified linear collision operator, which is similar to the one used in CODE \cite{landreman_numerical_2014}. The Abraham Lorentz-Dirac force due to the synchrotron radiation energy loss for the particle gyromotion in a magnetic field is included, and has been shown to alter the effective critical electric field for runaway generation and form vortex structures in electron momentum space. For RE avalanche, a Chiu-Harvey source term is included which gives a distribution of secondary REs depending on the energy distribution of seed REs.

Note that in previous studies, it is often assumed that the pitch angles of runaway electrons are very small due to the strong inductive electric field ($E_\parallel$) and weak Coulomb collisions. Given that these small pitch-angle electrons are strongly passing, the toroidal effect due to the inhomogeneity is not important. However, it has been demonstrated that high-$Z$ impurities, although weakly ionized, can cause strong pitch angle scattering of high energy REs through partial screening effect\cite{hesslow_effect_2017,breizman_physics_2019}. This effect can be included as a new effective Coulumnb logarithm in the collision operator\cite{hesslow_effect_2018}. Calculation based on density function theory (DFT) shows that for plasma with Ar$^+$, the slowing-down rate of runaway electrons can increase by a factor of 15, and the pitch angle scattering rate can increase by a factor of 150. We have modified the collision operator used in QUADRE to include both effects.

The strong pitch angle scattering from high-$Z$ impurities can cause the REs to have $p_\perp$ comparable to $p_\parallel$ and become trapped, where $p_\perp$ and $p_\parallel$ are momentum perpendicular and parallel to the local magnetic field. When trapped, the electron will not run away further because of the cancellation between the acceleration and deceleration of  $E_\parallel$ in forward and backward motion, and will loose energy due to collisions and radiation damping. In previous studies it has been shown that the trapping of REs can reduce the avalanche growth rate\cite{nilsson_trapped-electron_2015}. In order to take into account the bounce motion of runaway electrons, we have modified the QUADRE code to use a bounce-average kinetic equation which is similar to that in Ref.~\cite{rosenbluth_theory_1997}. The kinetic equation can be written as,
\begin{equation}
\label{eq:dfdt}
\frac{\partial f}{\partial t}\int \frac{|\xi|d\theta/2\pi}{\sqrt{1-\lambda b(\theta)}}=E[f]+C[f],
\end{equation}
\begin{align}
E[f]&=-\frac{eE}{mc}\left(\frac{1}{p^2}\frac{\partial}{\partial p}p^2 \xi f+\frac{1}{p^2}\frac{\partial}{\partial \xi}(1-\xi^2)p f\right)\int \sigma \frac{d\theta}{2\pi},\\
C[f]&=\frac{\ln\Lambda_1}{\ln\Lambda}\frac{1}{p^2}\frac{\partial}{\partial p}p^2\int \frac{|\xi|d\theta/2\pi}{\sqrt{1-\lambda b(\theta)}}\left(C_A\frac{\partial f}{\partial p}+C_F f\right)+\frac{\ln\Lambda_2}{\ln\Lambda}\frac{C_B}{p^2}\frac{\partial}{\partial \xi}\int \frac{\sqrt{1-\lambda b(\theta)}d\theta/2\pi}{|\xi|}\frac{\partial f}{\partial \xi},
\end{align}
where $f=f(p,\xi)$ is the electron distribution function in momentum space. $p=\gamma v/c$ is the relativistic momentum normalized to $mc$. $\xi$ and $\lambda$ represent the particle pitch angle at the minimum $B$ field along its orbit, which satisfy 
\begin{equation}
\lambda=1-\xi^2=\frac{p_\perp^2}{p^2}\frac{1}{b(\theta)},\qquad b(\theta)=\frac{B(\theta)}{B_{\mathrm{min}}}=\frac{1+\alpha}{1+\alpha\cos\theta},
\end{equation}
where $\theta$ is the poloidal angle. Here we assume $B(\theta)$ is inversely proportional to local major radius $R+r\cos\theta$, and $\alpha=r/R$ is the inverse of aspect ratio at the orbit location.  The integral of $\theta$ represents averaging along the particle orbit,  where we take the zero-orbit-width (ZOW) approximation and assume electrons stay on one flux surface. For passing electrons, $\theta$ is integrated from 0 to $2\pi$, and for trapped electrons $\theta$ is integrated from $-\theta_{max}$ to $\theta_{max}$ and back again where $\theta_{max}$ is the poloidal angle of banana orbit tip. The result of these terms can be written in the form of elliptic integrals,
\begin{equation}
\label{eq:integral-K}
\int \frac{d\theta/2\pi}{\sqrt{1-\lambda b(\theta)}}=\frac{2}{\pi\sqrt{1-\lambda}}\mathrm{Re} \left[K\left(\sqrt{\frac{2\alpha\lambda}{(1-\alpha)(1-\lambda)}}\right)\right],
\end{equation}
\begin{equation}
\int \sqrt{1-\lambda b(\theta)}d\theta/2\pi=\frac{2}{\pi}\sqrt{1-\lambda}\mathrm{Re}\left[ E\left(\sqrt{\frac{2\alpha\lambda}{(1-\alpha)(1-\lambda)}}\right)\right],
\end{equation}
where $K(\cdots)$ and $E(\cdots)$ are the complete elliptic integral of the first and the second kind.

In this set of equations, $E[f]$ is the electric field force, in which $E$ is the parallel electric field. $\sigma$ is +1 or -1 depending on whether the electron parallel motion is in the same or the opposite direction of the $E$ field. For passing electrons, the integral $\int \sigma d\theta/2\pi$ just gives 1 or -1 depending on the sign of $\xi$, whereas for trapped electrons this integral gives zero. $C[f]$ is the linearized collision electron operator including the slowing-down and pitch angle scattering, in which $C_A$, $C_F$ and $C_B$ are the collision coefficients introduced in Ref.~\cite{landreman_numerical_2014}. Note that this collision operator  can behave like a non-relativistic collision operator in the low energy regime, which is more complicated than that in Ref.~\cite{rosenbluth_theory_1997}. $\ln\Lambda$ is the Coulomb logarithm for runaway electrons, and $\ln\Lambda_1$ and $\ln\Lambda_2$ are the corrected logarithm factors due to the partial screening of impurities, which are discussed in Ref.~\cite{hesslow_effect_2018}. Note that in addition to these two operators, the synchrotron radiation damping force operator\cite{stahl_effective_2015} and Chiu-Harvey source term\cite{chiu_fokker-planck_1998} have also been included in the model after gyro-averaging. However, the synchrotron radiation damping is not as important as the collision term, given that most of the REs in the current study are below 10MeV and the radiation damping force is proportional to the relativistic factor $\gamma$. The Chiu-Harvey source term is also a subdominant term given that most of the REs in the DIII-D experiments are generated through hot-tail mechanism, because the inductive electric field is close to the critical electric field in the DIII-D disruption scenario and the avalanche is weak.

The kinetic equation is solved using a finite element method. Note that for bounce-averaged kinetic equation in the ZOW approximation, the continuity of the particle flux across the passing-trapped boundary needs to be treated carefully. The reason is that in the trapped region, $f$ is symmetric with respect to $\xi=0$ since the number of particles moving forward and backward are the same. Therefore, the derivative of $f$ is not necessarily continuous across the two passing-trapped boundaries. Instead, the particle flux into and out of the trapped region should be equal. This can be taken care of in the framework of the finite element method, by defining the Galerkin integral of the basis function at the passing-trapped boundary to be a combination of two integrals at the two sides of the boundary. In other words, if $v$ is a test function corresponding to a node at the passing-trapped boundary, then its integral in the Galerkin method is defined as
\begin{equation}
\begin{aligned}
&\left.\int p^2 dp d\xi \nu \frac{\partial f}{\partial t}\int \frac{|\xi|d\theta/2\pi}{\sqrt{1-\lambda b(\theta)}}\right\vert_{\xi=\xi_0}+\left.\int p^2 dp d\xi \nu \frac{\partial f}{\partial t}\int \frac{|\xi|d\theta/2\pi}{\sqrt{1-\lambda b(\theta)}}\right\vert_{\xi=-\xi_0}\\
&=\left.\int p^2 dp d\xi \nu \left[E[f]+C[f]\right]\right\vert_{\xi=\xi_0}+\left.\int p^2 dp \nu \left[E[f]+C[f]\right]\right\vert_{\xi=-\xi_0}
\end{aligned}
\end{equation} 
where $\xi_0=\sqrt{2\alpha/(1+\alpha)}$ is the value of $\xi$ at the passing-trapped boundary. Here $p^2 dp d\xi \nu \int |\xi|d\theta/2\pi/\sqrt{1-\lambda b(\theta)}$ is used as the Jacobian for the phase space volume. By combining the integral, the total number of equations in the Galerkin method will be reduced, which can be supplemented by forcing the value of $f$ at two boundaries to be equal.

\begin{figure}[h]
	\begin{center}
		\includegraphics[width=0.45\linewidth]{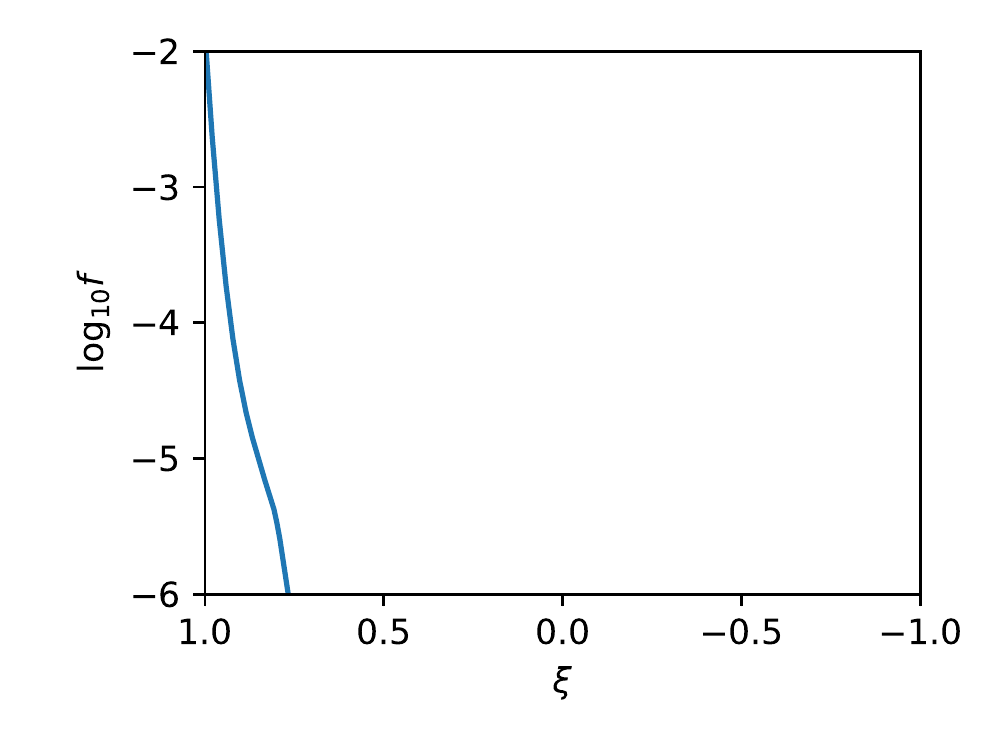}
		\includegraphics[width=0.45\linewidth]{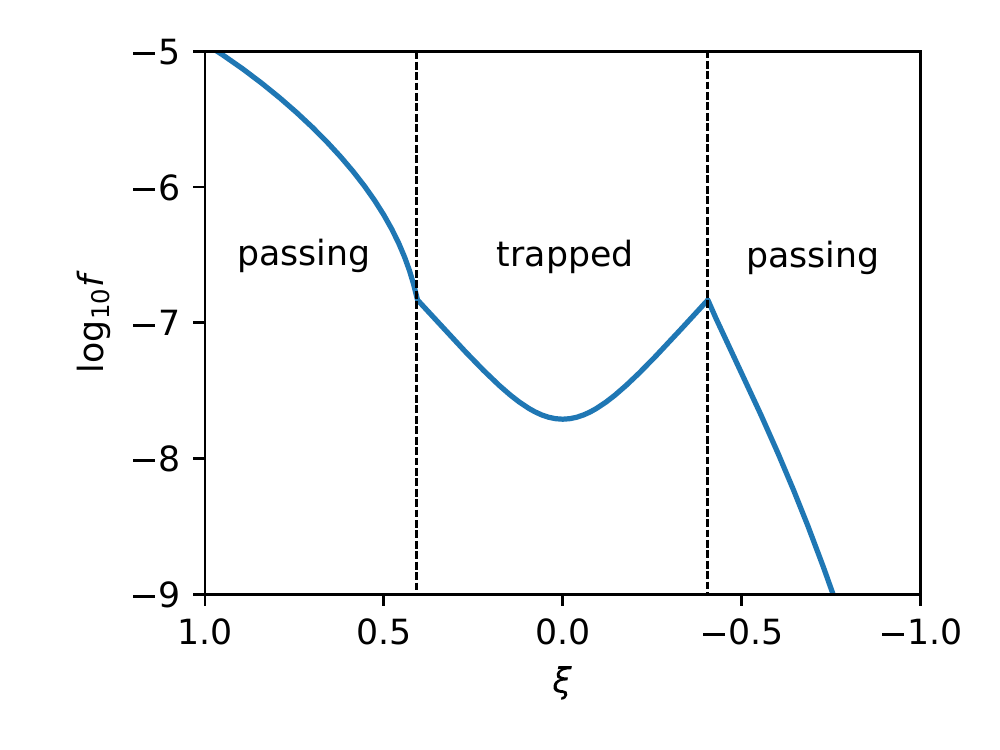}
	\end{center}
	\caption{\label{fig:f-pitch}Pitch angle distribution of $f$ for $p=7$. The left plot shows the result for deuterium plasma, and the right plot shows the result for plasma with Ar$^{2+}$. The dashed lines show the passing-trapped boundary.}
\end{figure}

The effect of the partial screening collision operator can be illustrated with a test simulation, and the result is shown in Fig.~\ref{fig:f-pitch}. In this simulation, the runaway electrons are generated through hot-tail mechanism due to a high electric field, and then accelerated to high energy with a fixed electric field about 3V/m. The left plot shows the resultant pitch angle distribution function at $p=7$ with deuterium and no impurities, and the right plot shows the result with Ar$^{2+}$. The aspect ratio $R/r=7$ is used to calculate $b(\theta)$, and the dashed line shows the location of trapped-passing boundary. We can see that with the enhanced pitch angle scattering due to the presence of Ar$^{2+}$, a  larger portion of generated REs are scattered into the large pitch angle region and become barely passing or trapped electrons compared to the deuterium case. The distribution inside the trapped region is symmetric with respect to $\xi=0$, and its derivative is not continuous at the passing-trapped boundary.

\section{Linear mode growth rate with runaway electrons}
\label{sec:wave-particle}

Similar to the whistler waves, the interaction between runaway electrons and Alfv\'en modes requires the electrons to satisfy the resonance condition. For REs there are 3 characteristic frequencies, the gyrofrequency, the transit (for passing) or bounce (for trapped) frequency, and the precession frequency. However, given that the frequencies of Alfv\'en modes have $\omega\ll \omega_{ce}$, the Doppler resonance condition discussed in the whistler wave interaction is not possible. Instead, the runaway electrons can satisfy the Cherenkov resonance, which is similar to the resonance condition satisfied by energetic ions for the low frequency shear Alfv\'en modes.

For REs moving in tokamaks, the Cherenkov resonance condition can be written as
\begin{equation}
\omega=n\omega_\phi-m\omega_\theta,
\end{equation}
where $\omega_\phi$ and $\omega_\theta$ are the electron transit frequencies in the toroidal and poloidal directions. For passing REs, $\omega_\theta\approx\omega_\phi/q$ if ignoring the drift motion and assuming electrons are following the magnetic fields.
Note that for strongly passing REs with small pitch angle, the transit frequency ($\sim$10~MHz) is still too large to match the resonance condition. Thus only the barely passing REs can satisfy the resonance condition with CAEs. On the other hand, for trapped REs, $\omega_\theta=0$ and $\omega_\phi$ is equal to the precession frequency $\omega_{d}$. For a trapped runaway electron with $E=5$~MeV in an orbit with minor radius $r=0.2$~m, major radius $R=1.4$~m, magnetic field $B=2$~T and safety factor $q=1$, the precession frequency $\omega_{d}\approx 1.45$MHz, which is of the same order of CAE frequencies, thus the resonance condition can be satisfied. Note that this resonance condition for trapped REs has also been used to explain the excitation of beta-induced Alfv\'en eigenmodes (BAE) observed in HL-2A experiments\cite{hl-2a_team_$ensuremathbeta$-induced_2010}.

For runaway electrons satisfying the resonance condition, the wave-particle energy exchange can be calculated as\cite{belikov_destabilization_2003,belikov_stabilization_2004}
\begin{equation}
G=\mathbf{E}\cdot\mathbf{v}_d+E_\perp v_\perp J_1(k_\perp \rho)
\end{equation}
where $\mathbf{E}$ is the electric field from the CAE and $E_\perp$ is the component perpendicular to both the equilibrium magnetic field and the direction of $\mathbf{k}_\perp$, $\mathbf{v}_d$ is the electron drift velocity, $v_\perp$ is the electron velocity perpendicular to the equilibrium magnetic field, and $\rho$ is the Larmor radius. The first term is the power of electric force acting on the drift motion of electrons. The second term is the energy exchange due to the gyromotion in a nonhomogeneous electric field, which will be reflected in the particle energy associate with electron magnetic moment $\mu B$. The second term is missing in the shear Alfv\'en wave analysis since in that case $\mathbf{E} \parallel \mathbf{k}_\perp$. This term reflects the energy exchange due to the compressional magnetic perturbation ($\delta B_\parallel$). Note that this term is responsible for the transit time magnetic pumping (TTMP) effect for the wave particle interaction\cite{stix_waves_1992}. Assuming $k_\perp \rho\ll 1$, the ratio of the two terms is about $(v_\perp/v)^2/k_\perp R$. Thus for runaway electrons with large pitch angles, the second term dominates the interaction with CAE fields. 

Using the above equations, the growth rate for the CAE with REs can be calculated as
\begin{equation}
\label{eq:gammaL}
\gamma_L=\frac{4\pi^2 e^{2}}{\mathcal{E}}\int \frac{|\langle G\rangle|^{2}}{\omega}\delta(\omega-n\omega_\phi+m\omega_{\theta})\left(\omega \frac{\partial}{\partial E}+n\frac{\partial}{\partial P_{\phi}}\right)_{\mu}f d^{3}\mathbf{p}
\end{equation}
where $E$ is the electron energy, $P_{\phi}=p_{\parallel}R-\psi$ is the electron toroidal momentum. $\langle\cdots\rangle$ is the average along the orbit. $\mathcal{E}$ is the total energy associated with the mode.

The derivative of $f$ in Eq.~(\ref{eq:gammaL}) illustrates that the mode will exchange both energy and toroidal momentum with the electron population with a fixed ratio, which is determined by the mode dispersion relation. This derivative can be written using variables $p_\parallel, p_\perp$ and $\psi$,
\begin{equation}
\label{eq:derivative}
(\omega \frac{\partial}{\partial E}+n\frac{\partial}{\partial P_{\phi}})_{\mu}=\frac{\omega}{v_{\parallel}}\left(\frac{\partial}{\partial p_{\parallel}}\right)_{p_{\perp},\psi}+\left(\frac{\omega R}{v_{\parallel}}-n\right)\left(\frac{\partial}{\partial \psi}\right)_{p_{\perp},p_{\parallel}}
\end{equation}
where we ignore the difference of $|B|$ at different $\psi$. This equation shows that a positive $\partial f/\partial p_\parallel$ (bump on tail) or a negative $\partial f/\partial \psi$ (peaked profile) can lead to a positive mode growth rate.

Note that in a rigorous calculation of the mode growth rate, both $\mathcal{E}$ and the integral in Eq.~(\ref{eq:gammaL}) should be calculated in the whole region of plasma utilizing the mode structure calculated in Sec.~\ref{sec:mode-structure}. This will require a 3D kinetic simulation for $f(\psi, p, \xi)$. Given that our current kinetic simulation is only 2D, we will use a plane wave approximation and only take the integral in the momentum space. The spatial gradient of distribution function $\partial f/\partial \psi$ comes from a  provided profile function of $f$. This allows us to get a quick solution for the mode growth rate, but the global effect of RE profile and the transport effect will be absent. The further development of kinetic simulation for multiple flux surfaces and a more rigorous calculation of $\gamma_L$ is left for future work.

\section{Collisional damping of compressional Alfv\'en eigenmodes}
\label{sec:collisional-damping}

In a post-disruption plasma with electron temperature only a few eV, the Alfv\'en modes will be susceptible to strong damping from electron collisions, which can be comparable to the drive from REs. This is similar to the collisional damping of the whistler waves\cite{aleynikov_stability_2015}, which plays a dominant role of determining the stability of modes. Note that this is different from the damping mechanism due to collisions of resonant trapped electrons which plays an important role in Toroidal Alfv\'en Eigenmodes (TAEs) in hot plasma\cite{fu_stability_1993}. For CAEs in disruptive plasma, the resonance can happen with runaway electrons which are almost collisionless. 

In Ref.~\cite{aleynikov_stability_2015} the collisional damping of whistler waves in disruptive plasma is calculated by adding a anti-Hermitian term into the dielectric tensor. This is equivalent to replacing the electron mass $m_e$ with $m_e(\omega+i\nu_{ei})/\omega$ in the dielectric tensor, where $\nu_{ei}$ is the electron-ion collision frequency. This method can be used to calculate the electron dominant mode like whistler wave with $\omega\gg \nu_{ei}$, by assuming the ions are fixed and electrons are colliding with them. However, for low frequency modes ($\omega\ll\omega_{ci}$) like CAE, the ion and electron motions are strongly correlated, and ignoring the ion motion can overestimate the collisional damping.

Here we show a more rigorous calculation of collisional damping by adding friction terms into the two-fluid equations. We study the CAE in a slab geometry with a background magnetic field in the $z$ direction (ignoring spatial inhomogeneities). By replacing the time derivative with $-i\omega$, the two fluid equations can be written as
\begin{equation}
\label{eq:friction}
\begin{aligned}
-i\omega V_{ix}&=eZ_{i}E_{x}/m_{i}+\omega_{ci} V_{iy}+\nu_{ie}(V_{ex}-V_{ix}),\\
-i\omega V_{iy}&=eZ_{i}E_{y}/m_{i}-\omega_{ci} V_{ix}+\nu_{ie}(V_{ey}-V_{iy}),\\
-i\omega V_{ex}&=-eE_{x}/m_{e}-\omega_{ce} V_{ey}+\nu_{ei}(V_{ix}-V_{ex}),\\
-i\omega V_{ey}&=-eE_{y}/m_{e}+\omega_{ce} V_{ex}+\nu_{ei}(V_{iy}-V_{ey}),
\end{aligned}
\end{equation}
where $V_{ix}$ and $V_{iy}$ are ion velocities in $x$ and $y$ direction, and $V_{ex}$ and $V_{ey}$ are the electron velocities. $E_x$ and $E_y$ are the electric fields. $m_i$ and $m_e$ are the ion and electron mass, and $\omega_{ci}$ and $\omega_{ce}$ are the ion and electron cyclotron frequency, and $\nu_{ie}$ is the ion-electron collision frequency. According to the momentum conservation law, the two collision frequencies satisfy $n_e m_e \nu_{ei}=n_i m_i \nu_{ie}$.

Given the electric fields, the velocities of the two fluid components can be calculated by solving the linear equations,
\begin{equation}
\label{eq:matrix-inversion}
\left(\begin{matrix}
V_{ix}\\
V_{iy}\\
V_{ex}\\
V_{ey}
\end{matrix}\right)=
\left(\begin{matrix}
i\omega-\nu_{ie}&\omega_{ci}&\nu_{ie}&0\\
-\omega_{ci}&i\omega-\nu_{ie}&0&\nu_{ie}\\
\nu_{ei}&0&i\omega-\nu_{ei}&-\omega_{ce}\\
0&\nu_{ei}&\omega_{ce}&i\omega-\nu_{ei}
\end{matrix}\right)^{-1}
\left(\begin{matrix}
-eZ_{i}E_{x}/m_{i}\\
-eZ_{i}E_{y}/m_{i}\\
eE_{x}/m_{e}\\
eE_{y}/m_{e}
\end{matrix}\right),
\end{equation}
and the plasma current can be calculated as
\begin{equation}
J_{x,y}=n_i V_{ix,y}-n_e V_{ex,y}.
\end{equation}
The current can be used to calculate the dielectric tensor in $x,y$ directions, which can help derive the dispersion relation of CAE. If we remove the collisional drag force by setting $\nu_{ei}=\nu_{ie}=0$, then this calculation just gives the standard dielectric tensor in magnetized plasma as in Ref.~\cite{stix_waves_1992}.

With $\nu_{ei}\ne 0$, the inversion of the matrix in Eq.~(\ref{eq:matrix-inversion}) becomes much more complicated to calculate, and the result cannot be written in a simple form. Assuming that the $\nu_{ei}\ll \omega_{ci}, \omega_{pi}$, we can solve the linear matrix order by order to find the solution of $\omega$, whose imaginary part is the collisional damping rate ($\gamma_d=-\mathrm{Im}\left(\omega\right)$). For CAE with $\omega\ll\omega_{ci}$, we only need the $yy$ component of the dielectric tensor. By keeping the lowest order correction of $\nu_{ei}$, the dispersion relation can be written as
\begin{equation}
  \frac{k^2 c^2}{\omega^2}=\epsilon_{yy}=1+\frac{\omega_{pi}^2}{\omega_{ci}^2}+\frac{\omega_{pe}^2}{\omega_{ce}^2}-\frac{\omega_{pi}^2 \left(\omega_{ci}+i\omega_{ce}\right)^2}{\omega_{ci}^3\omega_{ce}^3}\omega\nu_{ei},
\end{equation}
where $\omega_{pe}$ and $\omega_{pi}$ are the electron and ion plasma frequencies. Note that $\omega_{pi}^2/\omega_{ci}^2+\omega_{pe}^2/\omega_{ce}^2\approx c^2/v_A^2\gg 1$. Assuming that $\mathrm{Im}\left(\omega\right)\ll \mathrm{Re}\left(\omega\right)$, the damping rate can be solved perturbatively by substituting $\omega^2=k^2 v_A^2$,
\begin{equation}
\label{eq:gammad}
\gamma_d=\frac{ \left(\omega_{ci}+\omega_{ce}\right)^2}{2\omega_{ci}\omega_{ce}^3}\nu_{ei}\omega^2\approx \frac{\omega^2}{2\omega_{ci}\omega_{ce}}\nu_{ei},
\end{equation} 
where we used $\omega_{ci}\ll\omega_{ce}$.

\begin{figure}[h]
	\begin{center}
		\includegraphics[width=0.5\linewidth]{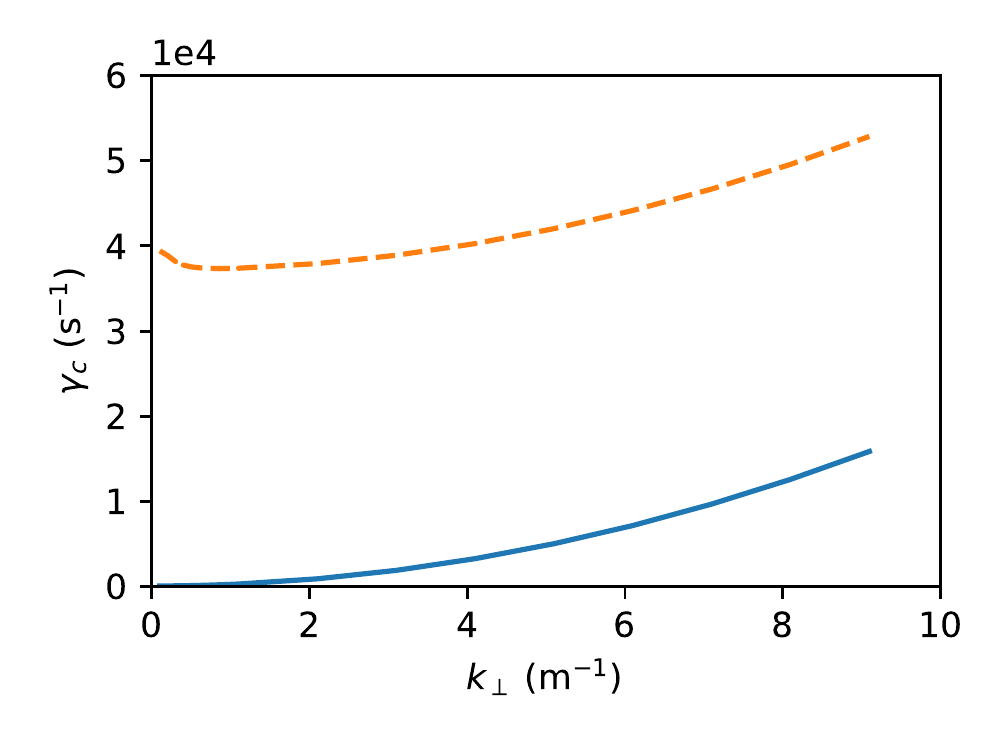}
	\end{center}
	\caption{\label{fig:collision-rate}Collisional damping rate of CAE calculated using Eq.~(\ref{eq:gammad}) (solid line), and the damping rate calculated using the method in Ref.~\cite{aleynikov_stability_2015} (dashed line).}
\end{figure}

The collisional damping rate obtained from Eq.~(\ref{eq:gammad}) is a quadratic function of $\omega$ and $k$. Note that if we use the method in Ref.~\cite{aleynikov_stability_2015} by assuming that the ions are fixed, we will get a nonzero $\gamma_d$ value at $\omega=0$. This is shown in Fig.~\ref{fig:collision-rate}, where $\gamma_d$ for CAE is calculated for $k_\parallel=0.7143$~m$^{-1}$ and different values of $k_\perp$. The plasma parameters are $n_e=2\times 10^{20}$~m$^{-3}$, $B=2.0$~T, $Z=2$, $m_i=40~m_H$ where $m_H$ is the hydrogen mass, and $T_e=5$~eV. The solid lines shows the result of $\gamma_d$ using Eq.~(\ref{eq:gammad}). The dashed line shows the result of $\gamma_d$ assuming the ions are fixed, which has a significant overestimate of the collisional damping for small $\omega$.

Given $\omega\ll\omega_{ci}$, Eq.~(\ref{eq:gammad}) shows that the damping rate can be much smaller comparable to $\nu_{ei}$ for CAE. However, for $\omega$ close to $\omega_{ci}$, the higher order terms become more important, and the collisional damping of the mode will increase significantly. This is consistent with the two-fluid analysis. As shown in Eq.~(\ref{eq:friction}), the collision damping comes from the collisional friction which is proportional to the velocity difference between electrons and ions. As CAE frequency increases, the separation between electron and ion motion becomes more significant, which can lead to a stronger damping effect due to the collision friction force.

\section{Linear simulation of compressional Alfv\'en eigenmodes in disruption scenarios}
\label{sec:linear-simulation}

In this section we show the result of a kinetic simulation of runaway electron dynamics and a calculation of CAE growth rates. The simulation is based on DIII-D shot 177028\cite{lvovskiy_role_2018}, in which a significant amount of REs are generated and Alfv\'en modes are thus observed. The kinetic simulation is based on the bounce-averaged model described in Sec.~\ref{sec:kinetic-model}, and the linear mode growth rates are calculated based on the Eq.~(\ref{eq:gammaL}). Note in the current version of the code, the kinetic simulation is only conducted on a single flux surface with a fixed value of $q$. Here we choose the flux surface at $R=1.4$~m, $r=0.2$~m with $q=1.08$. 

The initial electron distribution in the kinetic simulation is a Maxwellian with temperature $Te=1.2$~keV. After the simulation begins, a linearized collision operator with temperature 5eV is applied, to simulate the sudden cooling of the bulk electrons in thermal quench. In the first 20$\mu$s, a large parallel electric field is applied to drive the hot-tail generation of REs. After the generation, the smaller electric field ($E=3$~V/m) is applied, which is close to the inductive electric field measured at the tokamak edge in experiments. Plasma is composed by Ar$^{2+}$ and electrons with density $n_e=2\times 10^{20}$~m$^{-3}$. In this setup, the number of REs generated is about $n_{RE}\approx 5\times 10^{-4} n_e$, which compose a current about 1.17MA. Note that, for simplicity, the electric field in our simulation is fixed and not self-consistently determined because our focus is not on hot-tail generation history but rather on the evolution of the tail distribution function at later times.

\begin{figure}[h]
	\begin{center}
		\includegraphics[width=0.5\linewidth]{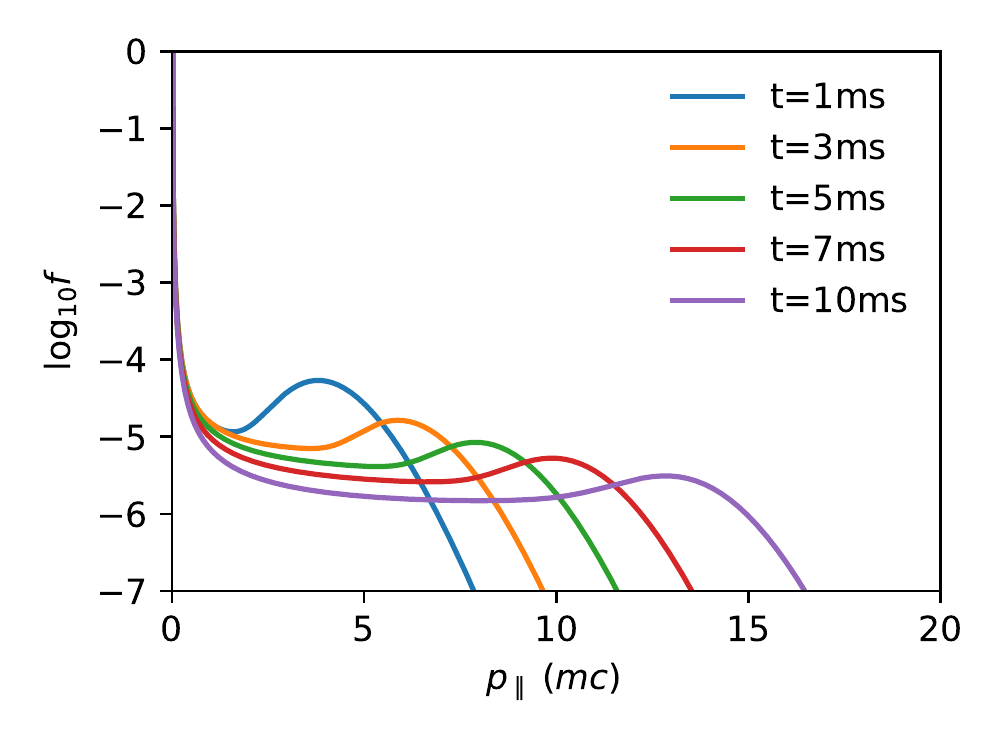}
	\end{center}
	\caption{\label{fig:fevolve}Evolution of RE distribution function at $f(p_\parallel, p_\perp)$ at $p_\perp=0$.}
\end{figure}

\begin{figure}[h]
	\begin{center}
		\includegraphics[width=0.32\linewidth]{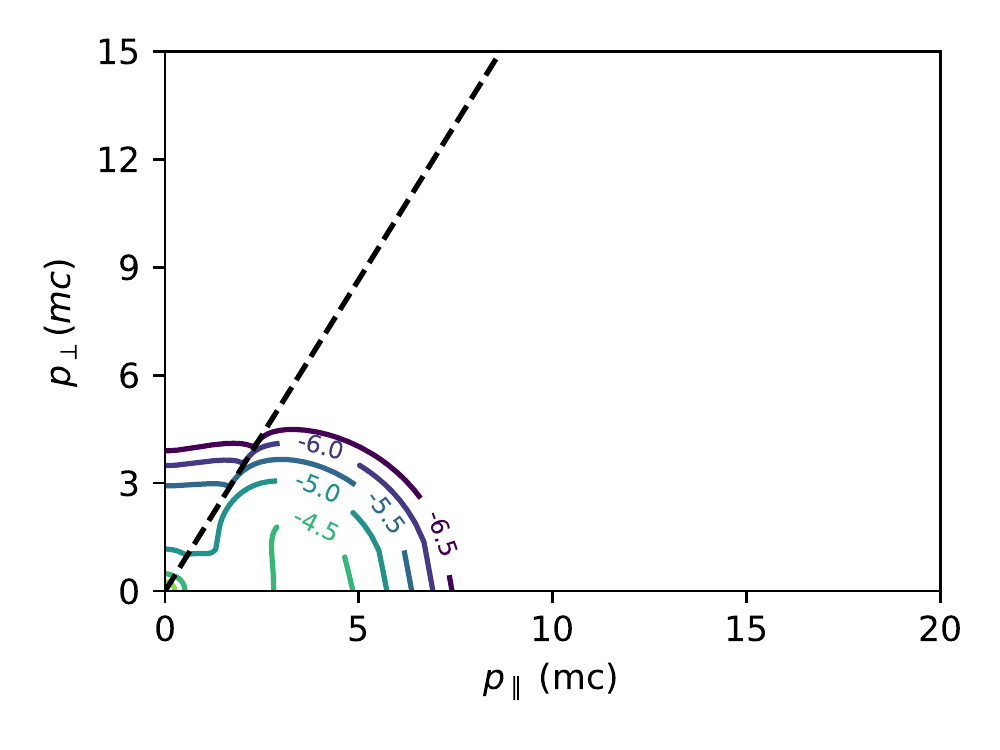}
		\includegraphics[width=0.32\linewidth]{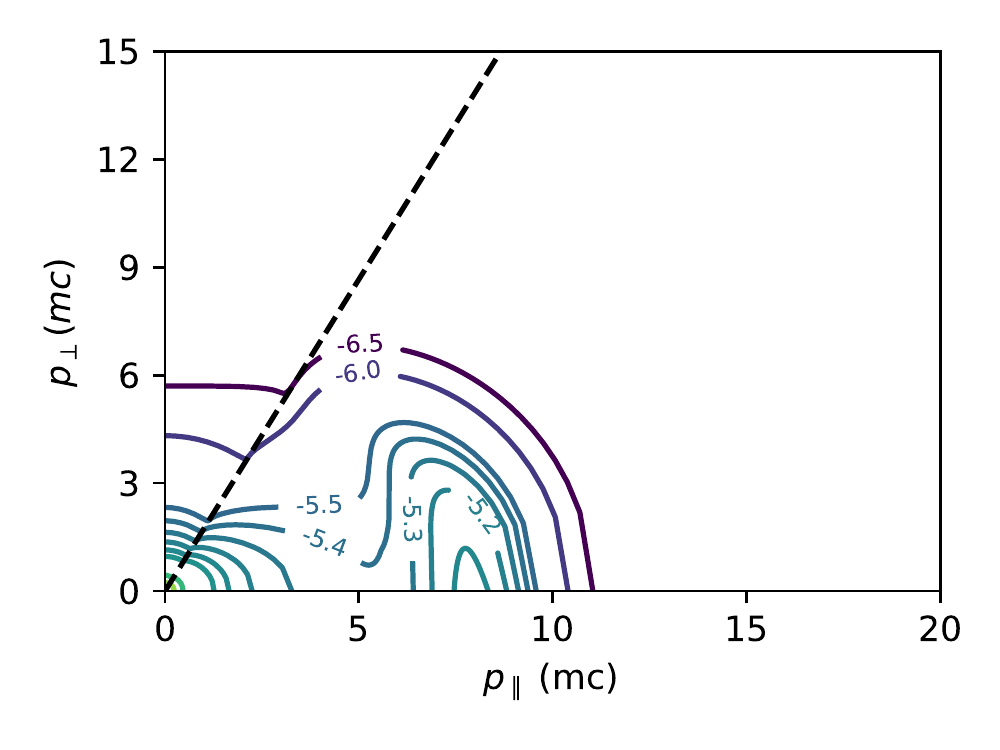}
		\includegraphics[width=0.32\linewidth]{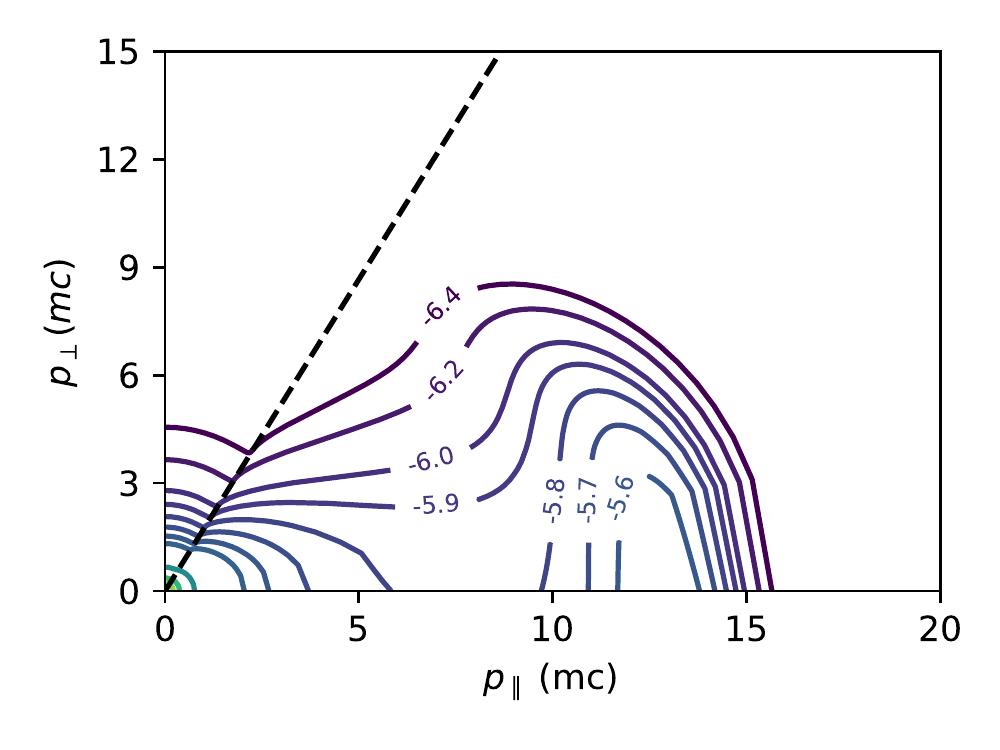}
	\end{center}
	\caption{\label{fig:f2d}Evolution of RE distribution function during the simulation. The three plots showing the contours of $\log_{10} f$ in $p_\parallel-p_\perp$ space at time $t=1$~ms, $t=5$~ms, and $t=10$~ms. The dashed lines show the passing-trapped boundaries.}
\end{figure}

The evolution of the electron distribution function is shown in Fig.~\ref{fig:fevolve} and Fig.~\ref{fig:f2d}. We can see that during the simulation, a RE tail is generated and dragged into the high energy regime, and a bump on tail distribution is formed. After this bump moves to high energy regime, a certain amount of REs are scattered to high pitch angle, and become barely passing and trapped electrons which are resonant with CAEs. In addition, the bump on tail ($\partial f/\partial p_\parallel>0$) can also come from the anisotropic distribution of REs ($\partial f/\partial \xi>0$) in the region of large $p_\perp$. This mechanism of forming bump-on-tail has been discussed in Ref.~\cite{liu_role_2018}, which can drive whistler waves in the 100MHz frequency range. 

Using the distribution function obtained from kinetic simulation, we can further calculate the CAE growth rate $\gamma=\gamma_L-\gamma_d$, where $\gamma_L$ is the growth rate from RE gradient using Eq.~(\ref{eq:gammaL}), and $\gamma_d$ is the collisional damping rate from Eq.~(\ref{eq:gammad}). As shown in Eq.~(\ref{eq:derivative}), $\gamma_L$ depends on the spatial gradient of $f$, which cannot be directly obtained from the kinetic simulation limited to one single flux surface. Nevertheless, the generated REs will have a peaked profile since hot tail generation depends sensitively on the plasma temperature before the thermal quench, and the trapped REs will have strong Ware pinch due to the parallel electric field. Here we assume that the electron distribution function satisfies an exponential profile with $f=\exp(-r^2/r_0^2)f(p,\xi)$ to calculate the spatial gradient, where $r_0=0.4$~m. The spatial gradient of $f$ at $r=0.2$~m can thus be modeled as $\partial f/\partial r=-f/r_0$.

\begin{figure}[h]
	\begin{center}
		\includegraphics[width=0.5\linewidth]{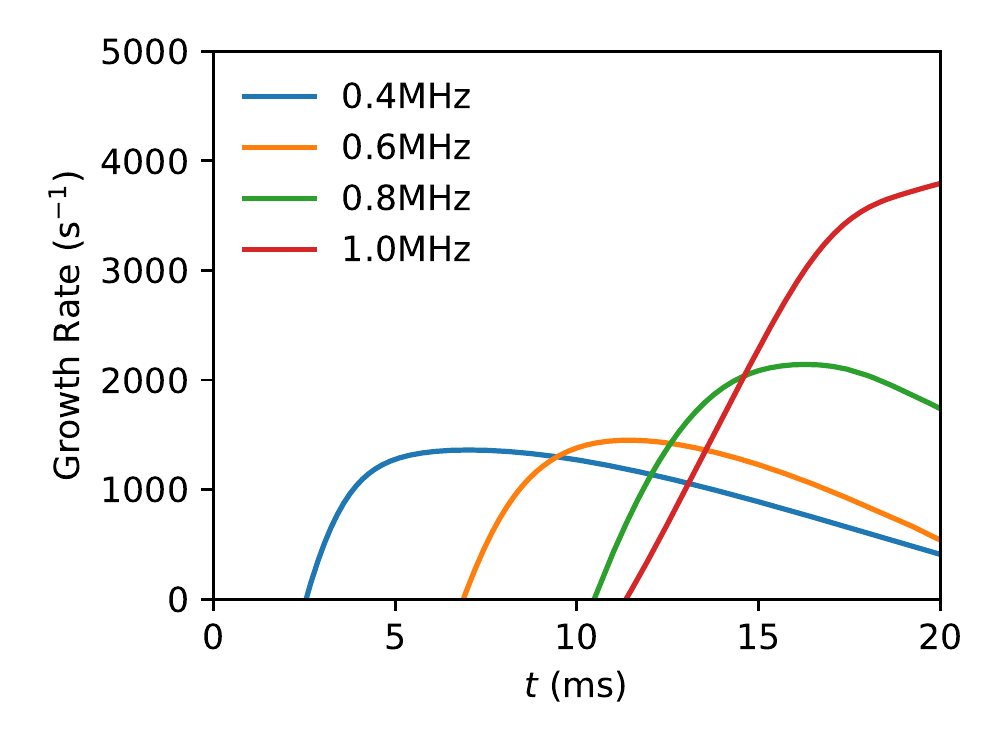}
	\end{center}
	\caption{\label{fig:growth-rate}Growth rate ($\gamma=\gamma_L-\gamma_d$) of $n=1$ CAEs during the kinetic simulation.}
\end{figure}

The result of $\gamma$ for CAE with $n=1$ is shown in Fig.~\ref{fig:growth-rate}. We can see that in a few milliseconds, the low frequency modes first become unstable, which is consistent with the experiments. This is because as REs are moving into the high energy regime, they first become resonant with the low frequency mode. In addition, these modes have smaller $\gamma_d$ and are easier to excite. After the bump of $f$ passes the resonance region, the growth rates of these modes will saturate and starts to drop, and higher frequency modes can become unstable. According to the linear calculation, the higher frequency modes can have larger growth rates, but this result is questionable because the excited low frequency mode can flatten the distribution function due to wave-particle interaction (WPI). To account for the behavior of the later-excited high frequency mode, a quasilinear or nonlinear simulation including WPI is required.

\begin{figure}[h]
	\begin{center}
		\includegraphics[width=0.5\linewidth]{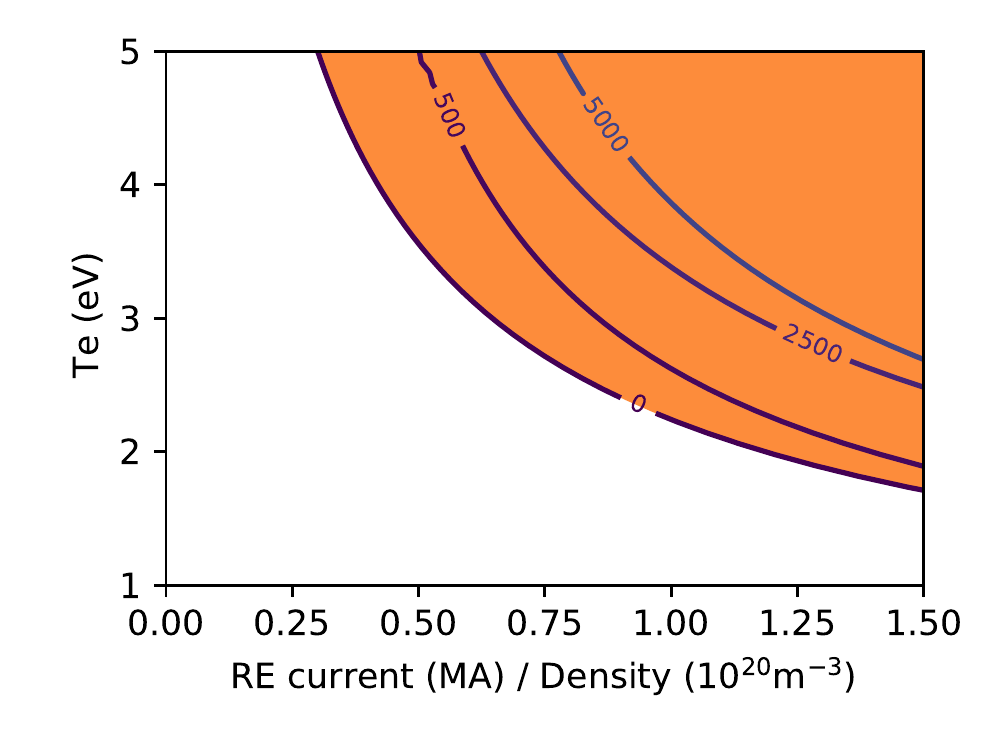}
	\end{center}
	\caption{\label{fig:stability}Contours of the maximum value of growth rate ($\gamma=\gamma_L-\gamma_d$) divided by $n_e$ ($10^{20}$m$^{-3}$) for $n=1$ CAEs in the DIII-D disruption scenario for different values of $I_{RE}$ and $T_e$. The orange region shows the parameter regime where the mode is unstable ($\gamma>0$). }
\end{figure}

We can use the result of $\gamma_L$ and $\gamma_d$ from the simulation to calculate the stability boundary of CAE. Note that $\gamma_L$ is proportional to the RE current $I_{RE}$, and $\gamma_d$ is proportional to $\nu_{ei}$ which depends on the electron density and temperature as $\nu_{ei}\sim n_e T_e^{-3/2}$. Therefore $\gamma$ for a certain mode at a specific time can be calculated as a function of $I_{RE}$, $n_e$ and $T_e$. Fig.~\ref{fig:stability} shows the maximum value of $\gamma$ divided by $n_e$ for $n=1$ modes during the simulation time for different values of $I_{RE}/n_e$ and $T_e$, which reveals a stability boundary ($\gamma=0$). Note that in this calculation it is assumed that argon is injected and becomes the dominant ion species, and the shape of the RE distribution function is unchanged. However, as plasma temperature drops below 2eV, the dominant charge state of argon change from +2 to +1, which can enhance the pitch angle scattering of REs due to partial screening effects. This can lead to the generation of more resonant runaway electrons, and thus make the CAEs more unstable.

\section{Summary and Discussion}
\label{sec:discussion}

In this paper we show observed instabilities in DIII-D post-disruption plasma can be explained by CAEs driven by barely passing and trapped runaway electrons, which can satisfy the resonance condition. The generation of these REs relies on collisions with partially ionized high-$Z$ impurities, which can produce enhanced pitch angle scattering due to partial screening effects. The resonant runaway electrons can drive the mode through the gradient of distribution function in momentum space (bump-on-tail) and in the radial direction, which is similar to the interaction of Alfv\'en eigenmodes with energetic ions. In addition, although in a plasma after thermal quench the collisional damping of the mode is strong, the damping rates are much smaller compared to the electron-ion collision frequency $\nu_{ei}$ for low frequency CAEs, thus the mode is easier to excite. The linear simulation clearly shows that CAEs can be excited by runaway electrons generated through hot-tail mechanism in a post-disruption plasma with electron temperature above 2eV and large RE current, with high-$Z$ impurities injection.

The stability and growth rates of CAEs are calculated based on a simplified kinetic simulation of REs. In deriving the bounced-average kinetic equation, a zero-orbit-width approximation is taken. In addition, only the distribution function at a single flux surface is used to estimate the mode growth rate. A more rigorous simulation can be done by extending the kinetic simulation from 2D to 3D, by adding the radial dimension and including the radial transport of REs. The dynamics of runaway electron tail can be better simulated using an inductive electric field calculated from $dI/dt$. The growth rate can then be calculated using the mode structure found in Sec.~\ref{sec:mode-structure} and integration among the flux surfaces. 

Note that another possibility for these high frequency modes is GAE, which has been studied recently through numerical simulation and used to explain the magnetic field perturbation observed in DIII-D experiments. However, However, the GAE is a type of shear-Alfven wave on the slow mode branch and its frequency is always below the ion cyclotron frequency ($\omega_{ci}$). For argon plasma, this frequency is about 2MHz, but the frequency spectrum observed in experiments appears to be higher than this upper limit. The dispersion equation of CAEs does not exhibit such a limit since it stays in the fast wave branch. The possibility of GAE excitation will be investigated in future study.

The excited modes will have feedback on the REs, which can lead to flattening of RE distribution at resonance lines. The resonance can form vortex structures in momentum space, which is similar to the wave particle interaction of whistler waves\cite{guo_phase-space_2017}. This will also result in a diffusion of resonant particles in the radial direction, and enhance the radial transport of REs. Moreover, if multiple CAEs are excited simultaneously, the resonance regions can overlap, and the diffusion can be enhanced like that for ions diffused by toroidal Alfv\'en eigenmodes (TAEs)\cite{pinches_observation_2006}. This enhanced diffusion of resonant REs gives a possible explanation for the fast dissipation of RE beam observed in DIII-D experiments\cite{lvovskiy_role_2018}.


The results above inspire us to explore another possible way to mitigate REs in the disruption, by utilizing RE diffusion effect of CAEs through launching the waves externally. As discussed in Sec.~\ref{sec:collisional-damping}, the CAE is less susceptible to collisional damping than whistler modes, so the power required to launch the mode externally is smaller. In addition, as discussed above, the mode can lead to diffusion of resonant REs in real space in addition to the  momentum space diffusion. The injection of high-$Z$ impurities can also help excite the mode by providing more trapped REs though pitch angle scattering. The results suggest that by judicious launching schemes of both types of waves, it may be possible to achieve suppression of RE avalanches as well as elimination of existing RE populations.

\ack
Chang Liu would like to thank Neal Crocker, Guo-Yong Fu, Håkan Smith, Nikolai Gorelenkov, Elena Belova, Jeff Lestz and Vinícius Duarte for fruitful discussion. This work was supported by the Simulation Center of electrons (SCREAM) SciDAC center by Office of Fusion Energy Science and Office of Advanced Scientific Computing of U. S. Department of Energy, under contract No. DE-SC0016268 and DE-AC02-09CH11466. This research used the high-performance computing cluster at Princeton Plasma Physics Laboratory and the Eddy cluster at Princeton University.

\section*{Disclaimer}

This report was prepared as an account of work sponsored by an agency of the United States Government. Neither the United States Government nor any agency thereof, nor any of their employees, makes any warranty, express or implied, or assumes any legal liability or responsibility for the accuracy, completeness, or usefulness of any information, apparatus, product, or process disclosed, or represents that its use would not infringe privately owned rights. Reference herein to any specific commercial product, process, or service by trade name, trademark, manufacturer, or otherwise, does not necessarily constitute or imply its endorsement, recommendation, or favoring by the United States Government or any agency thereof. The views and opinions of authors expressed herein do not necessarily state or reflect those of the United States Government or any agency thereof.

\section*{References}
	
\bibliography{paper}

\end{document}